\documentclass[11pt]{article}
\usepackage{amsfonts}
\newtheorem{theorem}{Theorem}
\newtheorem{definition}{Definition}

\begin{document}

\title{\bf Poisson integrators\/}
\author{
     B. Karas\"{o}zen\thanks{e-mail:
    bulent@metu.edu.tr}\\
       Department of Mathematics\\
         Middle East Technical University, 06531 Ankara-Turkey}
\maketitle

\abstract{An overview of Hamiltonian systems with noncanonical
Poisson structures is given. Examples of bi-Hamiltonian ode's,
pde's and lattice equations are presented. Numerical integrators
using generating functions, Hamiltonian splitting, symplectic
Runge-Kutta methods are discussed for Lie-Poisson systems and
Hamiltonian systems with a general Poisson structure.
Nambu-Poisson systems and the discrete gradient methods are also
presented.}\\


\section{Introduction}
In this review article we present Hamiltonian systems with a
noncanonical Poisson structure and give an overview of Poisson
structure preserving numerical integrators. In the last decade,
efficient and reliable numerical integrators were constructed for
canonical Hamiltonian systems with symplectic structure (see
\cite{hairer00gi, sanz-serna94nhp}). The theoretical foundation of
stability and convergence of symplectic integrators was provided
by the backward error analysis \cite{hairer00gi}. Methods based on
the preservation of the geometric features of the continuous
system by discretization are called "geometric integrators", which
include also the symplectic methods (see for a state of art review
of these methods and their applications \cite{budd01gia}). They
have better stability properties, smaller local and global errors
than the standard integrators based on the local error control.
The quadratic invariants of the underlying continuous system are
exactly preserved by symplectic integrators. Complicated
invariants, among them the polynomial invariants of order three
and greater are in general not preserved (\cite{budd01gia},
\cite{hairer00gi}). The phase space structure of low dimensional
Hamiltonian systems are very accurately preserved by symplectic
methods. But the applicability of symplectic integrators for
higher dimensional systems arising from semidiscretization of
Hamiltonian pde's is not well established \cite{ablowitz01dis}.

The noncanonical Hamiltonian systems on Poisson manifolds arise in
many applications. The best known example is the Euler top
equation, a three dimensional Hamiltonian system. The Poisson
structure is more general than the symplectic structure, the
dimension of a Poisson manifold doesn't need to be even as in the
case of the Euler top equation compared to symplectic manifolds
which appear only in even dimensions. The Poisson systems occur as
finite dimensional systems in the form of ode's and nonlinear
lattice equations or as infinite dimensional systems in the form
of pde's. Hamiltonian pde's like the Korteweg deVries (KdV)
equation, the nonlinear Schr\"odinger equation (NLS) and Euler
equations of an incompressible fluid have many applications in
hydrodynamics, optics and meteorology. The algebraic and geometric
properties of the Poisson manifolds, like the Poisson bracket,
Darboux transformations, Casimirs, symplectic foliation and
integrability, are given in Section 2. Many Hamiltonian systems
can be written with respect to the same coordinate system in more
than one Hamiltonian formulation. The so called bi-Hamiltonian
systems occur as ode's, pde's and in nonlinear lattices. The
properties of bi-Hamiltonian systems, and finite dimensional
examples  are given in Section 3. The real application of the
bi-Hamiltonian formulation is the field of Hamiltonian pde's and
nonlinear lattice equations, which are presented in Section 4.

Geometric integrators for noncanonical Hamiltonian systems or
Poisson systems are largely unexamined. There exist a variety of
methods for the Lie-Poisson systems, Hamiltonian systems with a
linear Poisson structure. The generating function approach
\cite{channell90sio} and splitting methods  based on the
separability of Hamiltonian for canonical Hamiltonian systems were
extended to the Lie-Poisson systems \cite{mclachlan93elp}.
Unfortunately there exits no general method for Hamiltonian
systems with a general Poisson structure beyond the Lie-Poisson
integrators. One can show that some of the symplectic methods  for
canonical Hamiltonian systems may preserve the Poisson structure
of some Hamiltonian systems. This means that whether an integrator
is Poisson or not, depends on the specific problem. For some
nonlinear lattice equations, it is possible by using the Darboux
coordinates to transform the Hamiltonian system with the Poisson
structure to a canonical one and then integrate this system by the
symplectic methods. But there exists no general algorithm for such
a transformation. These aspects of Poisson integrators are
discussed in Section 5.

In Section 6 the so called Nambu-Hamilton systems  which can be
considered as an extension of the Poisson structure are presented.
The structure of these systems is discussed and the discrete
gradient methods which applicable for them, are described.

\section{Poisson structures and Hamiltonian systems}

In order to present the finite dimensional and infinite
dimensional Hamiltonian systems, we have to look at the properties
of the Poisson manifolds and functions defined on these manifolds.
In our presentation we follow mainly \cite{olver93aol}, Chapters 6
\& 7. The key property here is the so called Poisson bracket,
which assigns to each pair of functions $F$ and $G$ in the space
of smooth functions  $\cal{F}({\cal M})$ on a smooth manifold
${\cal M}$ a third function,  denoted by $\{F,G\}$.
\begin{definition}The Poisson bracket is a bilinear operation satisfying
the following conditions:
\begin{itemize}
\item[(i)] skew-symmetry
$$\{F,G\} = - \{G,F\},$$
\item[(ii)] the Leibniz rule
$$\{F,G\cdot P\} = \{F,G\}\cdot P + G\cdot\{F,P\},$$
\item[(ii)] the Jacobi identity
$$\{\{F,G\},P\} + \{\{P,F\},G\} + \{\{G,P\},F\} = 0,$$
where $\cdot$ denotes the ordinary multiplication of functions.
\end{itemize}
\end{definition}
A manifold ${\cal M}$ with a Poisson bracket is called a Poisson
manifold and the bracket defines a Poisson structure on ${\cal
M}$. The Poisson manifold is more general then a symplectic
manifold; it doesn't need to be even dimensional. The canonical
Poisson bracket on a even dimensional Euclidean space ${\cal M} =
{\Bbb R}^{m},\;m =2n$ with the canonical coordinates
$(p,q)=(p_1,\ldots,p_n,q_1,\ldots,q_n)$ is given for two smooth
functions $F(p,q)$ and $G(p,q)$ by
\begin{equation}
\{F,G\} = \sum_{i=1}^n \left \{ \frac{\partial F}{\partial q_i}\frac{\partial G}{\partial p_i} -
\frac{\partial F}{\partial p_i}\frac{\partial G}{\partial q_i} \right ). \label{CPB}
\end{equation}
with the bracket identities
$$
\{p_i,p_j\} = 0,\qquad \{q_i,q_j\} =0,\qquad \{q_i,p_j\} =\delta_j^i,\;i,j=1,\ldots,n
$$
where $\delta_j^i$ denotes the Kronecker symbol.

The general local coordinate picture of a $m$ dimensional Poisson
manifold ${\cal M}$ is given by the Poisson bracket on local
coordinates $x=(x_1,\ldots,x_m)$. The Poisson bracket for two
smooth functions $F(x)$ and $G(x)$ takes the form
\begin{equation}
\{F,G\} = \sum_{i=1}^m \sum_{j=1}^m \{x_i,x_j\} \frac{\partial F}{\partial x_i}\frac{\partial G}{\partial x_j} \label{PB}
\end{equation}
A Hamiltonian vector field  $X_H$ on ${\cal M}$ with the
coefficients functions $\xi_i(x)$ depending on $H$ is given by
$$
X_H = \sum_{i=1}^m \xi_i(x) \frac{\partial}{\partial x_i}
$$
and satisfies
$$
X_H(F) = \{F,H\} = - \{H,F\}.
$$
The governing equations of  the flow associated with the vector
field are called Hamiltonian equations
\begin{equation}
\dot{x} = \{x,H\}. \label{HA}
\end{equation}
\begin{definition}
A general Hamiltonian system is given then by the triple ${\cal
M}$, the manifold, $\{\cdot,\cdot\}$, the Poisson structure and
$H(x)$  a real function on ${\cal M}$. The Poisson structure gives
the coordinate functions $J_{ij}= \{x_i,x_j\}$ which are called
structure functions. They satisfy the Leibniz rule and have the
following properties
\begin{itemize}
\item[i)] skew-symmetry:
$$J_{ij}(x) = -J_{ji}(x),\quad i,j=1,\ldots,m$$
\item[ii)] Jacobi identity:
\begin{equation}
\sum_{i=1}^m \{ J_{il}\partial_lJ_{jk} + J_{kl}\partial_lJ_{ij} +
J_{jl}\partial_lJ_{ki} \} = 0,\quad i,j=1,\ldots,m \label{JA}
\end{equation}
for all $x\in {\cal M}$ with $\partial_l = \partial/\partial x_l$.
\end{itemize}
\end{definition}
The conditions for the Jacobi identity (\ref{JA}) form a system of
nonlinear partial differential equations for the structure
functions $J_{ij}(x)$. The structure functions can be assembled
into a skew-symmetric structure matrix $J(x)\in {\Bbb R}^{m\times
m}$. In this case the Poisson bracket (\ref{PB}) for the functions
$F$ and $G$ takes the form
$$
\{ F,G\} = \nabla F(x)^T J(x) \nabla G(x)
$$
and the Hamiltonian equation (\ref{HA}) can be rewritten as
$$
\dot{x} = J(x) \nabla H(x).
$$
The most important examples of Poisson structures are associated
with $m$-dimensional Lie algebras. The associated Lie-Poisson
bracket between two functions $F,G$  with the structure constants
$c_{ij}^k\;,i,j,k=1,\ldots,m$ of the Lie algebra, is given by
$$
\{F,G\} = \sum_{j,k=1}^m c_{ij}^kx_k \frac{\partial F}{\partial
x_i}\frac{\partial G}{\partial x_j}. \label{LP}
$$
with the linear structure functions $J_{ij}(x) = \sum_{k=1}^m
c_{ij}^k x_k$.\\

\noindent{\bf Example 1: Euler top equation\/}\\ The most
important example of a Lie-Poisson system
$$
\frac{dx_i}{dt} = \sum_{i,j,k=1}^m c_{ij}^k x_k \frac{\partial H}{\partial x_j},\quad i=1,\ldots,m
$$
is the Euler top equation describing the motion of a rigid body
around a fixed point:
\begin{equation}
\dot{x}_1 =   \frac{I_2-I_3}{I_2I_3} x_2x_3, \quad
\dot{x}_2 =   \frac{I_3-I_1}{I_3I_1}  x_3x_1,\quad
\dot{x}_3 =  \frac{I_1-I_2}{I_1I_2} x_1x_2 \label{EUL}
\end{equation}
where $I_1,I_2,I_3$ denote the moments of inertia. The Poisson
structure is given by
\begin{equation}
\{x_1, x_2\} = -x_3,\quad \{x_2,x_3\}=-x_1, \quad \{x_3,x_1\}=-x_2
\label{EP1}
\end{equation}
It can be written in the Hamiltonian form (\ref{HA}) with the
energy integral $H$ as the Hamiltonian
\begin{equation}
H = \frac{1}{2} \left ( \frac{x_1^2}{I_1} + \frac{x_2^2}{I_2} + \frac{x_3^2}{I_3}\right ) \label{EULE}
\end{equation}
and with the linear structure matrix
$$
J(x) = \left (
\begin{array}{ccc}
0 & -x_3 & x_2 \\
x_3  & 0 & -x_1 \\
-x_2 & x_1 & 0
\end{array} \right ),
$$
$$
\frac{dx}{dt} = J(x) \nabla H(x) = x \times \nabla H(x)
$$
here $\times$ denotes the cross product.
\begin{theorem}A Poisson manifold of dimension $m$ is symplectic if the rank
of the Poisson structure is maximal everywhere. Because the rank
of a Poisson manifold at any point is always an even integer, a
symplectic manifold is  necessarily even dimensional. In other
words if the structure matrix $J(x)$ is invertible, then according
to Darboux's theorem the Poisson structure is isomorphic to the
standard symplectic structure in ${\Bbb R}^{2n}$. In this case the
structure matrix $J(x)$ determines a symplectic structure on
${\cal M}\subset {\Bbb R}^{m},\; m= 2n$ if and only if its inverse
$K(x) = J(x)^{-1}$ satisfies the following conditions
(\cite{olver93aol}, Proposition 6.15)
\begin{itemize}
\item[i)] skew-symmetry:
$$
K_{ij}(x) = -K_{ji}(x),\quad i,j=1,\ldots,m
$$
\item[ii)] Jacobi identity:
$$
\partial_k K_{ij}(x) + \partial_j K_{ki}(x) +\partial_i K_{jk}(x) = 0,\quad i,j,k=1,\ldots,m
$$
\end{itemize}
The nonlinear partial differential equations for the structure functions in the Jacobi identity (\ref{JA}) simplify to linear partial differential
equations involving $K_{ij}(x)$.
\end{theorem}
\begin{definition} \cite{olver93aol} A Poisson structure is called degenerate, if the structure
matrix $J(x)$is non-invertible. It follows from the skew-symmetry
that the rank of the Poisson matrix at any point is always an even
integer, which implies that odd dimensional systems have a
degenerate Poisson structure. If the manifold is of dimension
$m=2n+l$ and the rank of $J(x)$ is $2n$ everywhere, then there
exist distinguished functions $F_i(x) =c_i,i=1,\ldots,l$, called
Casimirs, whose Poisson bracket vanish with any function or
variable $x_i$ , i.e. $\{F_i(x),x_i\}=0,i=1,\ldots,l$.
\end{definition}
If the Poisson structure is of constant rank, the symplectic
foliation simplifies. One can introduce local coordinates, which
bring the foliation to the canonical form, which results from the
Darboux theorem (\cite{olver93aol}, Theorem 6.22). At each point
$x\in {\cal M}$ of an $m$-dimensional manifold of constant rank
$2n\le m$ everywhere, the Darboux theorem says then, there exist
local coordinates $(q_1,\ldots,q_n,p_1,\ldots,p_n,z_1,\ldots,z_l)$
in terms of which the Poisson bracket takes the canonical form
(\ref{CPB}), where $p_i,q_i$ denote the canonical variables. There
exist locally $l$ independent Casimirs $F_i(x),\cdots,F_l(x)$;
their common level surfaces $ F_1(x)=c_1,\cdots,F_l(x) =c_l $ are
called symplectic leaves of the Poisson bracket. Using the Darboux
transformation, the Poisson structure can be transformed on the
symplectic leaves to the canonical one.

If the structure matrix $J(x)$ is constant and nonsingular in some
coordinate systems, then a linear change of coordinates reduces
the Poisson structure to the canonical one \cite{dubrovin99fma}.
The reduction of the Poisson structure using Casimirs on
symplectic leaves produce usually more complicated systems. For
example for the Euler top equations, the total angular momentum
\begin{equation}
I=\frac{1}{2}(x_1^2+x_2^2+x_3^2), \label{EULM}
\end{equation}
is a Casimir (commutes with $x_i,i=1,\ldots,3$)
and the symplectic leaves are the spheres described by (\ref{EULM}).
Using the transformations
$$
q= \mbox{arccos} \left (\frac{x_1}{(x_1^2+x_2^2)^{1/2}}\right )
,\quad p=x_3
$$
the Euler top equation can be reduced to a canonical Hamiltonian system (\cite{blaszak98mto}, pp. 7)
$$
H(q,p,c) = (I-p^2) \left (\frac{1}{I_1}\cos ^2q + \frac{1}{I_2} \sin ^2 q \right ) + \frac{p^2}{I_3}.
$$

The geometric characterization of the local and global structure
of Poisson manifolds is given by the splitting theorem of Weinstein \cite{silva99gmf}:
\begin{theorem}
On a Poisson manifold ${\cal M}$, any point $x \in {\cal M}$ has a
coordinate neighborhood with coordinates
$q_1,\ldots,q_k,p_1,\ldots,p_k,y_1,\ldots,y_l$ centered at $x$,
such that
\begin{equation}
\prod_{}^{} = \sum_i \frac{\partial}{\partial q_i} \wedge
\frac{\partial}{\partial p_i} + \sum_{i,j} \varphi_{ij}(y)
\frac{\partial}{\partial y_i} \wedge \frac{\partial}{\partial
y_j}, \qquad \varphi_{ij}(0) = 0. \label{splitting}
\end{equation}
This theorem gives a decomposition of the neighborhood of $x$ as
the product of two Poisson manifolds: one with rank $2k$ and the
other with rank $0$ at $x$. The local structure of a Poisson
manifold is described by the symplectic leaves and their
transverse Poisson structures  (the second part in
\ref{splitting}). Globally,  a Poisson manifold is obtained by
gluing together the symplectic manifolds. There are three special
cases of the splitting theorem \cite{silva99gmf}:
\begin{itemize}
\item If the rank is locally constant, then $\varphi_{ij} =0$.
The splitting theorem reduces to the Lie's theorem, i.e. each
point of ${\cal M}$ is contained in a local coordinate system with
respect to which the structure functions $J_{ij}$ are constant
\item At the origin of ${\cal M}$ one has only $y_i$'s, the first term
in (\ref{splitting}) does not appear.
\item A symplectic manifold is a Poisson manifold where $rank
\;\prod = dim \;{\cal M}$ everywhere. The splitting theorem gives
canonical coordinates $q_1, \cdots,q_k,p_1,\cdots,p_k$.
\end{itemize}
\end{theorem}
There exists a close relation between the symmetries and
conservation laws of Hamiltonian systems. Noether's theorem
provides a connection between the symmetries and conservation laws
or first integrals and can be applied to Hamiltonian systems. The
first integrals are characterized by the vanishing of the Poisson
bracket: a function ${\cal I}$ is a first integral of the
autonomous Hamiltonian system (\ref{HA}), if $\{{\cal I}, H\}=0$
for all $x\in {\cal M}$. A Hamiltonian system with a
time-independent Hamiltonian has $H(x)$ as the first integral. The
Casimirs $F_i(x)$ determined by the Poisson structure $J(x)$  are
first integrals too, but they arise from the degeneracies of the
Poisson bracket and are not generated by the symmetry properties
of the Hamiltonian system. If the Poisson manifold is symplectic,
only constant functions are Casimirs.

Hamiltonian systems are usually non-integrable, there are in
general no additional integrals of motion. But there is a class of
Hamiltonians which are called completely integrable.
\begin{definition}\cite{rauch90hsa}
A Hamiltonian system is called completely integrable if it has $n$
integrals of motion $I_k(q,p), k=1,\ldots,n$
\begin{itemize}
\item which do not depend explicitly on time,
\item are functionally independent, i.e.
$$
{\rm rank } \left ( \frac{\partial I_k}{\partial
q_j};\frac{\partial I_k}{\partial p_j} \right )_{k,j=1,\ldots,n} =
n,
$$
\item are in involution
$$
\{ I_j,I_k\} = 0 \quad {\rm for\; each\; pair }\; (j,k),\quad
j,k=1,\ldots,n.
$$
\end{itemize}
\end{definition}
The Liouville theorem then says that any integrable system can be
integrated by quadratures. The geometric content of complete
integrability is described by the Liouville-Arnold theorem:
\begin{theorem}\cite{dubrovin99fma, perelomov95sto,rauch90hsa}
If the surface of the constant value of the Hamiltonian $H(q,p)=E$
is compact connected manifold, then
\begin{itemize}
\item the manifold
$$ {\cal M\/} = \{ (q,p): I_k(q,p) = c_k= {\rm const},\; k=1,\ldots,n\}
$$
diffeomorphic to the torus $T^n$,
\item in the canonically conjugate action and angle coordinates
$(I_k, \varphi_k)$, Hamilton equations have a simple form:
$$
\dot{\varphi}_i=\omega_k(I), \quad \dot{I}_k=0,\quad k=1,\ldots,n,
$$
\item the action-angle variables  have a simple dependence on time
$$
I_k(t)=const,\qquad \varphi_k(t) = \varphi_k^{(0)} + \omega_kt,
$$
the variable $\varphi_k$ determines the position of a point on the
torus and vary linearly with time; the motion is quasiperiodic on
the torus $T^n$,
\item the Hamilton equations can be solved by quadratures.
\end{itemize}
\end{theorem}
The Euler top equation is completely integrable. Other examples of
completely integrable  rigid body equations are the Lagrange top,
and the Kowalevskaya top equations \cite{arnold98tmi}.

If a Hamiltonian system has $2n-1$ independent  integrals at most, it is called completely degenerate. Such systems occur in very rare cases for
Hamiltonian with a potential function $V(q)=\alpha/q$ for the Kepler problem, $V(q)=\frac{1}{2}q^2$ for the harmonic oscillator and for the Calogero-Moser
system \cite{rauch90hsa}.

\section{Bi-Hamiltonian systems}

\begin{definition}A system of differential equations is called bi-Hamiltonian
if it can be written in the Hamiltonian form in two distinct ways
\cite{olver93cff} :
\begin{equation}
\frac{dx}{dt} = J_0(x) \nabla H_1 = J_1(x) \nabla H_0  \label{BIHA}
\end{equation}
with the structure matrices $J_0(x)$ and $J_1(x)$ determining the
Poisson  bracket $\{F,G\}_k = \nabla F^T J_k (x) \nabla G,\;
k=0,1$. The structure defined by $J_0(x)$ and $J_1(x)$ is called a
Hamiltonian pair. \end{definition}
A Hamiltonian pair is
compatible, if the linear combination of $J_1(x) - \lambda J_0(x)$
determines also a Poisson bracket for any real $\lambda$. The
compatibility condition is not trivial, because the differential
equations must be satisfied for the Jacobi identity (\ref{JA}). In
the symplectic case (non-degenerate $J_k (x),k=1,2$ and $n$ is
even), the Jacobi conditions can be replaced  by the symplectic
two-forms
$$
\Omega_k = \frac{1}{2} dx\wedge K_k(x) dx,\quad K_k(x) =
J_k^{-1}(x),\quad k=0,1
$$
which are closed, i.e.  $d\Omega_k = 0$. For a given pair, the
corresponding Hamiltonian system can be found by solving the
linear system of partial differential equations $\nabla H_1 =
J_1^{-1}(x)J_0(x)\nabla H_0$ \cite{olver90cfa}. A Hamiltonian pair
is called non-degenerate at the point $x$ if the skew-symmetric
matrix pencil $J_{\lambda }(x) = J_1(x) - \lambda J_0(x)$ is
nonsingular for at least one finite $\lambda$. According to Magri
\cite{magri96eli} bi-Hamiltonian systems with a compatible
Hamiltonian pair are completely integrable. \begin{theorem} For
any bi-Hamiltonian system with a non-degenerate, compatible
bi-Hamiltonian structure, there exists a hierarchy of Hamiltonian
functions $H_0,H_1,H_2,\ldots$, all in involution w.r.t. both
Poisson brackets $\{H_i,H_j\}_k =0,k=1,2,\;i,j=0,\ldots $ and
generating mutually commuting bi-Hamiltonian flows
(\cite{magri96eli, olver93cff}), i.e. they are completely
integrable. The sequence of first integrals can be constructed for
bi-Hamiltonian systems by the Lenard recursion scheme
(\cite{arnold98tmi,blaszak98mto, dubrovin99fma, magri96eli}):
\begin{eqnarray*}
\{ \cdot, H_0\}_0 & = & 0 \quad H_0 \mbox{ is a Casimir with respect to } \; \{\;,\;\}_0 \\
\{\cdot, H_1\}_0 & = & \{ \cdot, H_0\}_1, \\
\{\cdot, H_2\}_0 & = & \{ \cdot, H_1\}_1, \\
\cdots & \cdots & \cdots \\
\{\cdot, H_{k+1}\}_0 & = & \{ \cdot, H_k\}_1
\end{eqnarray*}
which generate a sequence of bi-Hamiltonian vector fields
\begin{eqnarray*}
X_k & = &  \{x,H_k\}_0,\\
& = & \{x,H_{k-1}\}_1,\quad k=1,2,\ldots
\end{eqnarray*}
\end{theorem}
In case of infinite dimensional systems (pde'e), one has an
infinite number of first integrals. For finite dimensional
systems, the Lenard recursion is either finite or infinite, in the
second case only a finite number of integrals are linearly
independent \cite{blaszak98mto}.\\

\noindent {\bf Example 2: Euler top equation as bi-Hamiltonian system\/}\\
The Euler top equation (\ref{EUL}) is also bi-Hamiltonian with
respect to the second Poisson bracket
\begin{equation}
\{x_1, x_2\} = \frac{x_3}{I_3},\quad \{x_2,x_3\}=\frac{x_1}{I_1}, \quad \{x_3,x_1\}=\frac{x_2}{I_2}\label{EP2}
\end{equation}
with the total angular momentum (\ref{EULM}) as the second Hamiltonian.
The corresponding structure matrix is given by
$$
J_1(x) = \left(
\begin{array}{ccc}
0 & \frac{x_3}{I_3} & -\frac{x_2}{I_2} \\
-\frac{x_3}{I_3} & 0 & \frac{x_1}{I_1} \\
\frac{x_2}{I_2} & -\frac{x_1}{I_1} & 0
\end{array}
\right).
$$

Examples of finite dimensional Hamiltonian systems with Poisson
and bi-Hamiltonian structures are given in \cite{gumral93pso},
\cite{perelomov95sto}.\\

\noindent {\bf Example 3: Lotka-Volterra equations\/}\\ Three
dimensional Lotka-Volterra system of competing species has a
bi-Hamiltonian structure
\begin{equation}
\dot{x} = \left ( \begin{array}{ccc}
0 & -x_1x_2 & x_1x_3 \\
x_1x_2 & 0 & -x_2x_3 \\
-x_1x_3 & x_2x_3 & 0
\end{array} \right ) \nabla H_1
=
\left ( \begin{array}{ccc}
0 & x_1x_2 x_3 & -x_1x_2x_3 \\
-x_1x_2x_3 & 0 & x_1x_2x_3 \\
x_1x_2x_3 & -x_1x_2x_3 & 0
\end{array} \right ) \nabla H_2 \label{VOL}
\end{equation}
with the Hamiltonians
\begin{equation}
H_1 = x_1 + x_2 + x_3\quad \mbox{and} \quad H_2 = \log (x_1) + \log (x_2) + \log (x_3). \label{VOLH}
\end{equation}
\noindent {\bf Example 4: Lorenz system\/}
\begin{eqnarray}
\dot{x_1} & = & \sigma (x_2-x_1) \nonumber \\
\dot{x_2} & = & (\rho x_1 - x_1x_3 - x_2)  \label{LOR} \\
\dot{x_3} & = & (-\beta x_3 +x_1x_2) \nonumber
\end{eqnarray}
has a chaotic behavior for most values of the parameters.
It is  Hamiltonian with the following parameter values and the Hamiltonian
$$
\rho = 0,\; \sigma =\frac{1}{2},\;\beta = 1,\quad
H= \frac{x_2^2+x_3^2}{(x_1^2-x_3)^2}.
$$
with respect to the Poisson structure
$$
\{x_1,x_2\} = x_1^2x_3+x_2^2, \quad \{x_2,x_3\} = -2(x_2^2+x_3^2)x_1,\quad \{x_3,x_1\} = (x_1^2-x_3)x_2.
$$
The system has two time-dependent conserved quantities.
Using  a transformation of the dynamical variables and time, the equation (\ref{LOR}) takes a bi-Hamiltonian form \cite{gumral93pso}
$$
u_1' =\frac{1}{2}u_2,\quad u_2'=-u_1u_3,\quad u_3'=u_1u_2.
$$
with the Hamiltonians $H_1 =u_3-u_1^2$, $H_2=u_2^2 + u_3^2$ and with the corresponding Poisson brackets
$$
\{u_1,u_2\}_0  =  \frac{1}{4},\quad \{u_2,u_3\}_0 =-\frac{1}{2}u_1,\quad \{u_1,u_3\}_0 = 0,
$$
$$
\{u_1,u_2\}_1  =  \frac{1}{2}u_3,\quad \{u_2,u_3\}_1 = 0,\quad \{u_1,u_3\} = \frac{1}{2}u_2
$$

Other three dimensional systems admitting bi-Hamiltonian structure
like the  Halphen system,  May Leonard equations, Maxwell-Bloch
equations, and the Kermack-McKendrick model for epidemics are
given in \cite{gumral93pso, perelomov95sto}.

There are few examples of bi-Hamiltonian systems in dimensions
higher than three \cite{blaszak98mto}. We give here as an example
the Calogero-Moser system with two particles.\\

\noindent {\bf Example 5: Calogero-Moser system\/}\\
The motion of repelling particles on a line is described by the
Hamiltonian
$$
H = \frac{1}{2}\sum_{j=1}^n p_j^2 + g^2 \sum_{j < k}
\frac{1}{(q_j-q_k)^2}.
$$
The four dimensional bi-Hamiltonian  Calogero-Moser system for two particles
is given in \cite{magri96eli} with  the Hamiltonians $H_0$ and $H_1$.
$$
H_1 =\frac{1}{2} (p_1^2 + p_2^2) + \frac{1}{(q_1-q_2)^2}, \quad H_0 = p_1 + p_2
$$
\begin{eqnarray*}
\dot{x} & = &
\left( \begin{array}{cccc}
0 &  0 & 1 & 0 \\
0 & 0 & 0 & 1 \\
-1 & 0 & 0 & 0\\
0 & -1 & 0 & 0
\end{array} \right) \nabla H_1 \\
& = &
\left( \begin{array}{cccc}
0 & \frac{2q_{12}}{4q_{12}^2 + p_{12}^2}  & p_1 + \frac{2q_{12}^2p_{12}}{4q_{12}^2 + p_{12}^2} & -\frac{2q_{12}^2}{4q_{12}^2 + p_{12}^2} \\
-\frac{2q_{12}}{4q_{12}^2 + p_{12}^2} & 0 &  0 & p_2 - \frac{2q_{12}^2p_{12}}{4q_{12}^2 + p_{12}^2} \\
-p_1 - \frac{2q_{12}^2p_{12}}{4q_{12}^2 + p_{12}^2}  & 0 & 0 & 2q_{12}^3 \\
\frac{2q_{12}}{4p_{12}^2 + p_{12}^2} & -p_2 + \frac{2q_{12}^2p_{12}}{4q_{12}^2 + p_{12}^2} & -2q_{12}^3 & 0
\end{array} \right) \nabla H_0
\end{eqnarray*}
where $q_{12} = \frac{1}{q_1 - q_2}$ and $p_{12}=p_1-p_2$.


\section{Bi-Hamiltonian pde's and lattice equations}
Many nonlinear partial differential equations like the Korteweg de
Vries (KdV) equation, the Euler equations of fluid dynamics and
the nonlinear Schr\"{o}dinger equation can be written in the
Hamiltonian form. The Hamiltonian formulation for the ode's can be
extended to pde's by replacing the Hamiltonian $H(x)$ by a
Hamiltonian functional ${\cal H}[u]$, the gradient operation
$\nabla H$ by the variational derivative $\delta {\cal H}$ and the
skew-symmetric matrix $J(x)$ by a skew-adjoint differential
operator $\Theta({\cal D})$. The resulting Hamiltonian system has
the form
$$
\frac{\partial u}{\partial t} = {\Theta(\cal  D})\times \delta {\cal H}[u]
$$
The right hand side of  the Hamiltonian evolution equations
$$
u_t = K[u]=K(x,u^{(n)})
$$
contains the space variable vector $x$ and $u^{(n)}$ represents
all derivatives of the function $u(x,t)$ with respect to $x$ of
order at most $n$. The Poisson bracket between two functional
${\cal F\/}$ and ${\cal G\/}$ is defined by
$$
\{ {\cal F},{\cal G}\} = \int \delta {\cal F} \cdot \Theta({\cal D}) {\cal G} dx
$$
which  satisfies the skew-symmetry condition and the Jacobi
identity. An interesting feature of the Hamiltonian pde's is that
many of them have a bi-Hamiltonian structure. As a result of this,
there exists a hierarchy of mutually commuting Hamiltonian flows,
and conserved quantities, generated by the recursion operator
based on two compatible Poisson operators. In the last two decades
the bi-Hamiltonian structures of many pde's were discovered, like
the KdV  equation, the nonlinear Schr\"{o}dinger equations, the
Euler equation, and the Boussinesq equation \cite{olver93aol}.

The Korteweg de Vries (KdV) equation has the bi-Hamiltonian form
\begin{eqnarray*}
u_t = u_{xxx} + uu_x & = & D_x\delta \int \left (-\frac{1}{2}u_x^2+\frac{1}{6}u^3\right ) dx=\Theta_0\delta H_1(u)  \\
 & = & \left(D_x^3 +\frac{2}{3}uD_x + \frac{1}{3}u_x\right ) \delta \int \frac{1}{2} u^2 dx = \Theta_1 \delta H_0(u)
\end{eqnarray*}

The nonlinear Schr\"odinger (NLS) equation with the complex
variable $u$
$$
iu_t +u_{xx} +2|u|^2u = 0, \label{NLS}
$$
has the following bi-Hamiltonian structure \cite{blaszak98mto}
\begin{eqnarray*}
\left( \begin{array}{c}
u_t\\
u^*_t
\end{array} \right) & = &
\left( \begin{array}{c}
iu_{xx} +2iu|u|^2\\
-iu^*_{xx} - 2iu^*|u|^2
\end{array} \right) \\
 & = &
\left(\begin{array}{cc}
0 & -i \\
i & 0
\end{array} \right) \delta \int_R (|u_x|^2 - |u|^4)dx
\end{eqnarray*}
$$
= \left(\begin{array}{cc}
2uD^{-1}u & -D-2uD^{-1}u^* \\
-D -2u^*D^{-1}u & 2u^*D^{-1}u^*
\end{array} \right) \delta \int_R \frac{1}{2}i(u^*_xu -u_xu^*)dx
$$
with $u^*$ denoting the complex conjugate of $u$. Both systems,
the KdV  and NLS equations, possess due to their bi-Hamiltonian
structure,  infinitely many first integrals, i.e. they are
completely integrable. The integrals can be generated using the
Lenard scheme. If the differential operators $\Theta_0$ and
$\Theta_1$ are compatible, the infinite hierarchy of
bi-Hamiltonian systems is defined recursively by
$$
u_t = \Theta_0 \delta H_{i+1}=\Theta_1\delta H_i, \quad i=-1,\ldots
$$
with the integro-differential  recursion operator $R=\Theta_1\Theta_0^{-1}$ \cite{antonowicz90hso}.\\

Nonlinear lattice equations or discrete nonlinear systems play an
important role in  modelling many physical phenomena. Examples of
them  are the Fermi-Pasta-Ulam lattice, the Toda and Volterra
lattices, the Calogero-Moser system. They also arise by
semi-discretization of some integrable nonlinear pde's like the
Ablowitz-Ladik lattice which represents an integrable
discretization of the nonlinear Schr\"{o}dinger equation
(\ref{NLS}). Most of the nonlinear lattices are completely
integrable. The conserved quantities, the multi-Hamiltonian
structure, the recursion operator, Lax form, and master symmetries
of the lattice equations were investigated in recent years by many
authors (see for example \cite{cherdantsev96lms, oevel89mam}).\\

\noindent{\bf Example 6: Toda lattice\/}\\ The motion of $n$
interacting particles with an exponential interaction potential
Hamiltonian
$$
H =  \sum_{i=1}^n \frac{1}{2}p_i^2 + e^{(q_i-q_{i+1}})
$$
are described by the Toda lattice equations. Here the $q_i's$
denote the positions of the $i$-th particle and the $p_i's$ their
momentum. Introducing the new variables $ a_i =
e^{(q_i-q_{i+1})},\; b_i = p_i $ after Flaschka transformation,
one obtains \cite{perelomov90iso, damianou00mhs}
\begin{equation}
\dot{a}_i = a_i(b_{i+1} - b_i),\;  \dot{b}_i = a_i - a_{i-1},\quad 1 \le i \le n \label{TOD}
\end{equation}
with the open-end  ($a_0=a_n=0)$ or periodic ($a_0=a_n,\;b_{n+1} =
b_1$) boundary conditions. It has a tri-Hamiltonian structure with
the following Poisson brackets and corresponding Hamiltonians
 \cite{suris99idf}
\begin{itemize}
\item linear Poisson bracket (Lie-Poisson bracket):
$$
\{b_i,a_i\}_1 = a_i, \quad \{a_i,b_{i+1}\}_1 = a_i
$$
$$
H_2 = \frac{1}{2} \sum_{i=1}^n b_i^2 + \sum_{i=1}^Na_i
$$
\item quadratic Poisson bracket:
$$
\begin{array}{ll}
\{b_i,a_i\}_2 = a_ib_i, & \quad \{a_i,b_{i+1}\}_2 = a_ib_{i+1} \\
\{a_i,a_{i+1}\}_2 = a_{i+1}a_i, & \quad \{b_i,b_{i+1}\}_2 = a_i \\
\end{array}
$$
$$
H_1 = \sum_{i=1}^n b_i
$$
\item cubic Poisson bracket:
$$
\begin{array}{ll}
\{b_i,a_i\}_3 = a_i(b_i^2+u_i), & \quad \{a_i,b_{i+1}\}_3 = a_i(b^2_{i+1} + a_i) \\
\{a_i,a_{i+1}\}_3 = 2a_ia_{i+1}b_{i+1}, & \quad \{b_i,b_{i+1}\}_3 = a_i(b_i +b_{i+1}) \\
\{a_i,b_{i+1}\}_3 = a_ia_{i+1}, & \quad \{b_i,a_{i+1}\}_3 = a_ia_{i+1}
\end{array}
$$
$$
H_1 = \frac{1}{2} \sum_{i=1}^n \log (a_i)
$$
\end{itemize}
With respect to the linear and quadratic Poisson brackets, the Toda lattice forms a bi-Hamiltonian system. Because both brackets are degenerate,
it is not possible to find the recursion operator and apply the Lenard scheme to find the hierarchy of the integrals, which was done using master symmetries
\cite{damianou00mhs}. A different bi-Hamiltonian formulation of the Toda can be found in \cite{damianou00mhs,oevel89mam}

Another interesting nonlinear lattice equation is the Volterra
lattice \cite{suris99idf} which is a model for vibrations of the
particles on lattices (Liouville model on the lattice) and
describes the population evolution in a hierarchical system of
competing species. It can  be also considered as an integrable
discretization of Korteweg de Vries equations.\\

\noindent{\bf Example 7: Volterra lattice\/}
\begin{equation}
\dot{y}_i = y_i(y_{i+1} - y_{i-1}), \quad i=1,\ldots,n \label{VOLN}
\end{equation}
 is bi-Hamiltonian
$$
\dot{y} =  J_0(y)
\nabla H_1
 = J_1(y)
\nabla H_0
$$
with respect to the quadratic and cubic Poisson brackets
\begin{equation}
\{ y_i,y_{i+1} \}_0 = y_iy_{i+1},
\end{equation}
\begin{equation}
\{ y_i,y_{i+1} \}_1 =  y_iy_{i+1} (y_i + y_{i+1}),\quad
\{ y_i,y_{i+2} \}_1 =  y_iy_{i+1} y_{i+2}.
\end{equation}
and the corresponding Hamiltonians are
$$
H_1= \sum_{i=1}^n y_i,\qquad
H_0 = \frac{1}{2}\sum_{i=1}^n \log (y_i).
$$
The Toda and Volterra lattices are closely related through the nonlinear
transformation of variables \cite{damianou00mhs}
$$
a_i = y_{2i}y_{2i-1},\qquad b_i =y_{2i-1} - y_{2i-2}.
$$

\section{Lie-Poisson and Poisson integrators}

It is natural to ask if the symplectic integrators applied to
canonical Hamiltonian systems on symplectic spaces can be extended
to produce integrators that preserve the linear Poisson structure
(Lie-Poisson integrators) or the more general Poisson structures.

\subsection{Integrators based on generating functions}
The generating function approach in the framework of symplectic
integrators was applied by several authors to Lie-Poisson systems.
Ge and Marsden constructed Lie-Poisson integrators in
\cite{ge-zhong88lph} by finding approximate solutions to the
Hamilton-Jacobi Lie-Poisson equations using the generating
functions developed in \cite{channell90sio} for canonical
Hamiltonian systems which is symplectic and energy preserving. The
time-dependent Hamilton-Jacobi equation is
$$
\frac{\partial S}{\partial t}  + H\left(q,\frac{\partial S}{\partial q}\right) = 0
$$
with $S(q,q_0,t)$. A canonical transformation
$\phi_S:(q_0,p_0)\rightarrow (q,p)$ is generated using
$$
p_0 = -\frac{\partial S}{\partial q_0},\quad p=\frac{\partial S}{\partial q}
$$
The initial conditions are chosen so that $S$ generates an
identity transformation at $t=0$
$$
S = \frac{1}{2t}(q-q_0)^2
$$
The generating function is expanded
$$
S = \sum_{i=0}^\infty \frac{\Delta t^n}{n!} S_i(p_0,q)
$$
and truncated to get an approximate solution $S_N$. The functions $S_i$ are computed recursively  and they satisfy the reversibility
condition for canonical  Hamiltonian systems with the Hamiltonian $H = \frac{1}{2}p^2 + V(q)$
$$
S_i(q,q_0,t) = -S_i(q_0,q,-t)
$$
The Ge-Marsden algorithm in \cite{ge-zhong88lph} is based on a
reduction technique which is in fact a symplectic integrator on
the symplectic leaves. The resulting scheme preserves the
Lie-Poisson structure on the reduced space, i.e. it is symplectic
on the symplectic leaves and preserves the angular momentum.
Ge-Marsden algorithm is an implicit scheme for general Lie-Poisson
systems, but it becomes explicit for the Euler top equations
(\ref{EUL}). The application of the Ge-Marsden algorithm for the
Euler top equations (\ref{EUL}) was also given in
\cite{ge-zhong88lph}. Similarly in \cite{ge90gfh} generating
functions are developed for general Poisson maps, which are
symplectic on the symplectic leaves. The Ge-Marsden algorithm is
then developed using exponential maps in the general Lie algebra
setting in \cite{channel91ifl} and applied to truncated finite
dimensional Vasto-Poisson equations. Numerical results show the
long term preservation of the energy and therefore the stability
of the algorithm. The Ge-Marsden algorithm is composed to obtain
higher order Lie-Poisson integrators in \cite{benzel93eco} and
applied again to Vasto-Poisson equations. A modification of the
Ge-Marsden algorithm is given in \cite{li95anf} and this is
illustrated for the Euler top equation (\ref{EUL}).

\subsection{Lie-Poisson integrators}
The implicit midpoint scheme preserves the constant Poisson
structure \cite{puta99cps}. It was shown in \cite{austin93api}
that the implicit midpoint scheme is an almost Poisson integrator,
that it preserves the Lie-Poisson structure up to the second order
terms of time step size $\Delta t$. The invariance of symplectic
difference schemes under symplectic transformations is discussed
in \cite{ge91esd}. The Hamiltonian function takes different forms
in different coordinates, so the symplectic schemes are coordinate
dependent, i.e.. they are not covariant. This can be shown for the
midpoint scheme, which is not invariant under general symplectic
coordinate transformations. Let the midpoint scheme  for the
canonical Hamiltonian system $\dot{u} = J^{-1}H_u$ be
\begin{equation}
\frac{u_{n+1} - u_n}{\Delta t} = J^{-1}H_u\left( \frac{1}{2}(u_n+ u_{n+1}\right) \label{MID}
\end{equation}
In an another coordinate system $v$, $u=S(v)$, where $S$ is a
symplectic coordinate transformation, the Hamiltonian becomes
$H(S(v))$and the midpoint scheme becomes
\begin{equation}
\frac{v_{n+1} - v_n}{\Delta t} = J^{-1}H_v\left( S(\frac{1}{2}(v_n+ v_{n+1})\right) \label{TMID}
\end{equation}
while the original midpoint scheme (\ref{MID}) in the coordinates $v$ is
\begin{equation}
\frac{S(v_{n+1}) - S(v_n)}{\Delta t} = J^{-1}H_v\left( \frac{1}{2}(S(v_n)+ S(v_{n+1})\right) \label{MIDT}
\end{equation}
If $S$ is nonlinear then the schemes (\ref{TMID}) and (\ref{MIDT})
are not identical. If $S$ is a linear symplectic transformation
$S^TJS=J$, then (\ref{TMID}) and (\ref{MIDT}) are identical. It
was shown that some symplectic difference schemes are invariant
under certain symplectic coordinate transformations and such
schemes are studied for Lie-Poisson systems in \cite{ge91esd}.

For Lie-Poisson systems in three dimensions, it is more convenient to work with the one form \cite{gumral93pso}. For the Euler top
equation the one form to be preserved becomes \cite{ergenc96rkc}
\begin{equation}
J = \sum_{i,j,k}=\epsilon_{ijk}J_{ij}dx_k = 2 \sum_{i=1}^3 x_idx_i \label{ONF}
\end{equation}
It was shown in \cite{ergenc96rkc} that (\ref{ONF}) is preserved
by the symplectic Runge-Kutta methods. Because the energy integral
(7) and the total angular momentum (8) of Euler top equation are
quadratic, they are preserved exactly by symplectic Runge-Kutta
methods of Gauss-Legendry family  and by partitioned Runge-Kutta
methods of Lobated ILIA-B type. A detailed  comparison of various
methods for the rigid body equations can be found in
\cite{buss00aae}.

Explicit Lie-Poisson integrators can be constructed, if the
Hamiltonian is separable as for the canonical Hamiltonian systems.
The linear structure of flows of Lie-Poisson systems were
exploited for the construction of explicit Lie-Poisson integrators
\cite{mclachlan93elp} and in \cite{reich94mcs} for the Euler top
equation, truncated Vasto-Poisson equations, and Euler equations
in fluid dynamics. If the Hamiltonian can be written as $H =
\sum_{i=1}^n H_i(x_i)$ , then each vector field $X_i = J\nabla
H_i$ can be integrated separately by an integrator $\phi_i(\Delta
t)$. In the case of the Euler top equation (\ref{EUL}) one
obtains,  with the Hamiltonians $ H_i = \frac{x_i^2}{2I_i}, \quad
i=1,\ldots, 3,$ the following splitting:
$$
\begin{array}{ccc}
\dot{x}_1 & = & 0 \\
\dot{x}_2 & = & \frac{x_1x_3}{I_1} \\
\dot{x}_3 & = &  -\frac{x_1x_2}{I_1}\\
\end{array},\quad
\begin{array}{ccc}
\dot{x}_1 & = & -\frac{x_2x_3}{I_2} \\
\dot{x}_2 & = &  0 \\
\dot{x}_3 & = &  \frac{x_1x_2}{I_2}\\
\end{array},\quad
\begin{array}{ccc}
\dot{x}_1 & = &  \frac{x_2x_3}{I_3} \\
\dot{x}_2 & = & -\frac{x_1x_3}{I_3} \\
\dot{x}_3 & = & 0 \\
\end{array}
$$ A composition
of the discrete flows
$$
\phi_1(\Delta t) \circ \cdots \circ \phi_n(\Delta t)
$$
gives a first order explicit integrator for the whole system. A
reversible second order composition can also be constructed
$$
\phi_1(\Delta t/2) \circ \cdots \circ \phi_n(\Delta t) \circ \cdots \circ \phi_1(\Delta t/2).
$$
The Casimirs, which are the angular momenta of the Euler top
equation, are preserved exactly. The Hamiltonian splitting
technique was applied to some Lie-Poisson systems, which are
obtained as finite dimensional models of some Hamiltonian pde's in
plasma physics and fluid dynamics. The truncated Vasto-Poisson
equations and the sine-Euler equations \cite{zeitlin91fao} as
finite dimensional approximations to the two-dimensional Euler
equation were integrated in \cite{mclachlan93elp} using the
Hamiltonian splitting technique and it was shown that all Casimirs
are preserved up to the round-off errors. It turns out that
explicit Lie-Poisson integrators are by far faster than the
Lie-Poisson integrators using generating functions in
\cite{channel91ifl, ge-zhong88lph}.

Time reversible integrators of arbitrary order were  constructed
in \cite{karasozen96cif} for some three dimensional bi-Hamiltonian
systems like Lotka-Volterra equations, Lorenz equation and for the
Toda lattice. Because the Hamiltonians of all these systems are
separable for both Poisson brackets, methods based on Hamiltonian
splitting can be applied. Symplectic composition of split flows
gives  an explicit second order time-reversible integrator. Higher
order composition techniques  \cite{yoshida90coh} are used to
increase the order of the methods in \cite{karasozen96cif}. The
Hamiltonians are preserved over long time intervals and the
periodicity of solutions were retained.

In \cite{feng91sds}  for nondependence Poisson systems with
$2n$-dimensional symplectic structure $K(x) = J^{-1}(x)$
symplectic difference schemes were constructed using a
parameterized nonlinear transformation from the Poisson manifold
to the canonical symplectic space. But as stated in
\cite{feng91sds}, in general it is difficult to find parameters
which preserve the Poisson structure.

In \cite{zhu94psf}, the integration of Poisson systems with
constant Poisson structure is considered and it was concluded that
from the class of symplectic Runge-Kutta methods, only the
diagonally implicit ones preserve the constant Poisson structure.
However Mclachlan pointed out in \cite{mclachlan95cop}, that
because all Runge-Kutta methods are equivalent under linear change
of coordinates, the constant  structure is also equivalent to a
canonical Poisson structure, therefore all symplectic Runge-Kutta
methods preserve the constant Poisson structure.

\subsection{Semidiscretized Hamiltonian pde's and integrable discretizations}
Nonlinear pde's in the Hamiltonian form can be discretized in
space, so that the resulting ode inherits the integrals of the
pde, this process  is known as the integrable discretization.
There are some examples of nonlinear pde's like the
Landau-Lifschitz equation and the NLS equation, which possess
integrable discretized nonlinear lattice equations.\\

\noindent{\bf Example 8: Landau Lifschitz equation}\\
The Landau-Lifschitz (LL) equation is a Lie-Poisson Hamiltonian
pde \cite{faddeev87hmi, frank97gif}
\begin{equation}
\frac{\partial S}{\partial t} = S \times \nabla^2S +  S \times DS
\label{LL}
\end{equation}
and was proposed originally as a model of an anisotropic
Heisenberg ferromagnet, where $S=S(x,y,t)\in {\Bbb
R}^3,\;||S||_2=1$ denote the spin length, the matrix $D\in {\Bbb
R}^{3\times 3}$ represents the anisotropy and may be assumed
diagonal with $|d_1|\le |d_2|\le |d_3|$.

The semi-discretized form of LL on a one-dimensional domain\
$\Omega = (-L,L)$ gives with $\{x_i\equiv-L+ih,\;i=0,\ldots,n\}$
the one-dimensional isotropic Heisenberg spin chain
\begin{equation}
\dot{S}_i = \frac{1}{h^2} S_i \times (S_{i-1} + S_{i+1}).
\label{LL1}
\end{equation}
The two-dimensional discretized equation over a rectangular domain
$\Omega = (-L,L) \times (-L, L)$ with $\{(x_i,y_j)\equiv
(-L+ih,-L+jh),i,j=0,\ldots,n\}$,  where $h=2L/n$ and $n$ a
positive even integer, has the form
\begin{equation}
\dot{S}_{ij} = S_{i,j}\times  M (S_{i,j-1} + S_{i,j+1} + S_{i-1,j} + S_{i+1,j}) \label{LL2}
\end{equation}
where
$$
S_{ij} \approx S(ih,jk,t) = \frac{1}{4} (S_{i,j-1} + S_{i,j+1} + S_{i-1,j} + S_{i+1,j})
$$
and $M=I/h^2 +D/4$. As boundary conditions periodic or homogenous
boundary conditions are taken.  The equations (\ref{LL1}) and
(\ref{LL2}) are Lie-Poisson systems and are referred to, for
$h=1$, as lattice Landau Lifschitz (LLL) equations. In two
dimensional case, the vector $S$ is defined by using the natural
ordering on the grid points:
$$
S = [ S^T_{1,1}, \ldots, S^T_{1,n}\; S^T_{2,1}, \ldots,
S^T_{2,n}\cdots S^T_{n,1}, \ldots, S^T_{n,n}]^T
$$
the equation (25) can be written as a Lie-Poisson system
$$
\dot{S} = J(S) \nabla_S H
$$
with the Hamiltonian
$$
H = -\sum_{i,j} S_{i,j} M (S_{i,j-1} + S_{i,j+1} + S_{i-1,j} +
S_{i+1,j}).
$$
We use the following notation to describe the form of the
structure matrix $J(S)$. For a vector $u\in {\Bbb R}^3$, we
associate a $3\times 3$ skew-symmetric matrix skew$(u)$ such that,
for any other vector $v\in {\Bbb R}^3$,  $v \times u \equiv {\rm
skew}(u)v$. Then the structure matrix $J(S)$ consists of blocks of
the  skew-symmetric matrices skew$(S_{i,j})$ on the diagonal. The
LL equations (24) and (25) have two additional integrals; the
individual spin lengths $||S_{ij}||_2$ and the total spin length
$\sum_i S_i$. This system was integrated for the one dimensional
case in \cite{frank97gif} by the splitting of the total
Hamiltonian $H=-\sum_i S_i \cdot S_{i+1}$ as $H=H_1 + H_2$
$$
H_1 = - \sum_i S_i\cdot S_{i+1}, \quad H_2 = - \sum_i S_i\cdot
S_{i-1}, \qquad \mbox{ for }  i \mbox { odd }
$$
The resulting Lie-Poisson systems
$$
\dot{S} = J(S) \nabla H_1,\qquad \dot{S} = J(S) \nabla H_2
$$
are then integrated with a Lie-Poisson integrator and the
corresponding flows are composed in a symmetric form. Another
discretization is based on the partitioning of the whole lattice
equation. In case of the one dimensional semidiscrete system the
spins are ordered as
$$
P_i \equiv S_{2i-1}, \quad Q_i \equiv S_{2i},\quad i=1,\ldots,n/2.
$$
We obtain then the following equations
$$
\dot{P}_i = P_i \times (Q_{i-1} + Q_i),\qquad \dot{Q}_i =Q_i
\times (P_i + P_{i+1}), \quad i=1,\ldots, n/2.
$$
Defining a vector filed splitting $V_H = V_1 + V_2$ by
$$
V_1 = \left [ \begin{array}{c} P_1 \times (Q_{n/2} + Q_1) \\
0\\
\vdots\\ P_i \times (Q_{i-1} + Q_i)\\
0\\
\vdots\\ P_{n/2} \times (Q_{(n-1)/2} + Q_{n/2})\\
0
\end{array}
\right ], \qquad V_2 =
\left [ \begin{array}{c} 0 \\Q_1 \times (P_1 + P_2) \\
\vdots\\0\\ Q_i \times (P_i + P_{i+1})\\
\vdots\\ 0\\ Q_{n/2} \times (P_{n/2} + P_1)
\end{array}
\right ]
$$
one can easily see that each of the vector fields are exactly
integrable. This system was integrated in \cite{frank97gif}  by a
semi-explicit staggered scheme which correspond to the second
order Lobatto IIIA-B pair \cite{hairer00gi}.

Both methods are time-reversible and preserve the individual
lengths of the spin vectors and total spin length. In terms of
operation counts the staggered scheme is cheaper than the
Lie-Poisson integrator, especially in two and three dimensions.\\

\noindent {\bf Example 9: Nonlinear Schr\"odinger equation}\\
The semi-discretization of the NLS equation (\ref{NLS}) by finite
differences or by spectral methods and the time integration of the
resulting integrable and nonintegrable ode's was the subject of
many papers in recent years. The standard finite discretization
scheme results in the direct NLS system (DNLS)
\cite{ablowitz96otn1}
\begin{equation}
i\dot{U}_j = -\frac{1}{h^2}(U_{j-1} -2U_j + U_{j+1}) -2|U_j|^2U_j \label{DNLS}
\end{equation}
where $h = L/N$ denote the grid spacing. DNLS  (\ref{DNLS})
possesses canonical Hamiltonian structure with the Hamiltonian
$$
H = -i\sum_{j=0}^{N-1} \left ( \frac{1}{h^2} |U_{j+1} - U_j|^2 - |U_j|^4\right)
$$
and with the only one additional integral of motion (the energy) $
I = \sum_{j=0}^{N-1} |U_j|^2 $. This system has only two integrals
whereas the NLS has infinitely many integrals, therefore the DNLS
(\ref{DNLS}) is not integrable. An integrable discretization of
(\ref{NLS}) is given by the Ablowitz Ladik  discrete NLS equations
(IDNLS):
\begin{equation}
i\dot{U}_j = -\frac{1}{h^2}(U_{j-1} -2U_j + U_{j+1}) -|U_j|^2(U_{j-1}+U_{j+1}) \label{IDNLS}
\end{equation}
with the Poisson bracket
$$
\{q_m,p_n\} = \frac{1}{h} (1+q_np_n)\delta_{m,n}, \quad \{q_m,q_n\} = \{p_m,p_n\}
$$
the Hamiltonian
$$
H = -\frac{i}{h^3}\sum_{j=0}^{N-1} \left ( h^2p_n(q_{n-1} + q_{n+1}) + 2 \log (1 +h^2 q_np_n )\right)
$$
and an additional first integral
$$
I = \sum_{j=0}^{N-1} q_j(p_{j-1} + p_{j+1})
$$
where $p_n = U_n$ and $q_n=U^*_n$. The equation (\ref{IDNLS}) has
$n$ integrals, i.e. it is completely integrable.  The
nonintegrable DNLS (\ref{DNLS}) was integrated in \cite{tang97smf}
by symplectic Runge-Kutta methods, which preserve the Hamiltonian
and the energy integral. The integrable discretization of NLS, the
IDNLS equations (\ref{IDNLS}) were transformed in the canonical
symplectic form using a Darboux transformation in
\cite{tang97smf}. The resulting system was then integrated with
the symplectic Runge-Kutta methods which preserve the Hamiltonian
and the integrals in long-term, whereas the nonsymplectic methods
fail.

The integrable discrete NLS equation (\ref{IDNLS}) was integrated
in \cite{schober99sif} by the construction of a generating
function to preserve the Poisson structure. A linear drift in the
Hamiltonian error was observed, and it was eliminated by a
modification of the scheme in \cite{hairer00sif}.

The following splitting of the Ablowitz-Ladik discrete NLS equations was introduced in \cite{suris97ano}:
$$
\begin{array}{cccc}
 & F^{-} \mbox{-flow } & F^0 \mbox{-flow } & F^{+} \mbox{-flow }\\
  & & & \\
 \dot{q_n} = & q_{n-1}(1-q_np_n) & -2q_n & + q_{n+1}(1-q_np_n) \\
 \dot{p_n} = & -p_{n+1}(1-q_np_n) & +2p_n & - p_{n-1}(1-q_np_n)
\end{array}
$$
The flows $F^{-}$ and $F^{+}$ are integrated by the symplectic
Euler method and  $F^0$ by the implicit midpoint rule and the
whole scheme is obtained by a symmetric decomposition. It was also
shown that the Poisson structure for each flow was preserved by
the symplectic Euler method \cite{suris97ano}. Unfortunately the
preservation of the Poisson structure for numerical methods is not
independent of the particular equations as in case of the
Hamiltonian systems with symplectic structure. A method which is
Poisson preserving for one system may not preserve the Poisson
structure of another system \cite{suris97ano}. The integration
above requires, unfortunately, very small step sizes
\cite{black99aon}, which makes it useless for numerical
computations.\\

\noindent {\bf Example 10: Volterra lattice equations:}\\
The Volterra lattice equations (\ref{VOLN}) can also be
represented with $u_i = y_{2i -1},\; v_i = y_{2i}$ in partitioned
form
\begin{equation}
\dot{u}_i  = u_i(v_i-v_{i-1}),\quad \dot{v}_i  =
v_i(u_{i+1}-u_i),\quad i=1,\ldots,m/2 \label{LVS}
\end{equation}
for even $m$ \cite{suris99idf}. The equations (\ref{LVS}) are also
bi-Hamiltonian with the quadratic and cubic Poisson brackets like
the original Volterra lattice equations. They were integrated with
the partitioned Lobatto IIIA-B methods in \cite{karasozen00spf}.
It was shown that the Poisson structure is preserved by the first
order symplectic Euler method and by the second order Lobatto
IIIA-B methods, whereas any of the Poisson brackets of the
Volterra lattice (\ref{VOLN}) can be preserved by the implicit
mid-point method.

Several geometric integrators for the NLS were compared recently
in \cite{ablowitz01dis}, \cite{islas01gif}, among them the
multisymplectic integrators. An alternative formulation of the
Hamiltonian pde's involves a local concept of symplecticness in
space and in time . Many Hamiltonian pde's are integrated using
finite difference and spectral methods by multisymplectic
integrators, which show better conservation properties than the
symplectic integration of semi-discretized pde's (see for example
\cite{reich00mrk}). An alternative formulation of
multisymplecticness and the variational integrators, which are
designed to integrate the Hamiltonian pde's are given in
\cite{mardsen99mgv}. These methods also have good preservation
properties. Which geometric integrators are the best for
Hamiltonian pde's, is still an open question.

\section{Nambu-Hamilton systems and discrete gradient methods}

In 1973 Nambu \cite{nambu73ghd} introduced a generalization of the
classical Hamiltonian mechanics and an extension of the Poisson
bracket. In  this formulation, the Poisson bracket is replaced by
a ternary or n-ary operation, called Nambu bracket. The underlying
idea of this new formulation was that in statistical mechanics the
basic result is the Liouville theorem, which follows from but does
not require the Hamilton dynamics.

\begin{definition}
In the
Euclidean space ${\Bbb R}^2$ with the coordinates $x_1$ and $x_2$
the Nambu bracket can be written as
$$
\{F_1,F_2\} = \frac{\partial F_1}{\partial x_1}\frac{\partial F_2}{\partial x_2}-\frac{\partial F_1}{\partial x_2}\frac{\partial F_2}{\partial x_1}=
\frac{\partial (F_1,F_2)}{\partial (x_1,x_2)}
$$
where this bracket satisfies the Jacobi identity (\ref{JA}). The
canonical Nambu bracket is defined for a triple of variables
$F_1,F_2,F_3$ in the phase space ${\Bbb R}^3$ with the coordinates
$x_1,x_2,x_3$ by the formula
$$
\{F_1,F_2,F_3\} = \frac{\partial(F_1,F_2,F_3)}{\partial(x_1,x_2,x_3)}
$$
where the right-hand side stands for the Jacobian.
\end{definition}
The first Nambu formulation was given for the Euler top equation
with the two Hamiltonians $H_1$, the energy integral (\ref{EULE})
and $H_2$, the total angular momentum (\ref{EULM}):
$$
\frac{dx}{dt} = \{H_1,H_2,x\} = \nabla H_1 \times \nabla H_2.
$$
The Liouville theorem is still valid; the corresponding phase flow
preserves the phase space volume. The orbits of a Nambu-Hamilton
system in the phase space are determined as the intersection of
the surfaces $H_1=$const and $H_2=$const. For the  Euler top
equation,  it is given by the intersection of the sphere described
by the total angular momentum $H_2$, and the total energy $H_1$.
The trajectories can be found by Jacobi elliptic functions. Nambu
introduced also a $n-$dimensional generalization of the Euler top
equation \cite{nambu73ghd}:
\begin{eqnarray*}
\frac{dx_i}{dt} & = &  \sum_{jk\cdots l} \epsilon_{ijk\cdots l}
\frac{\partial H_1}{\partial x_j} \frac{\partial H_k}{\partial
x_k}\cdots
\frac{\partial H_{n-1}}{\partial x_k},\\
\frac{dx}{dt} & = &  \frac{\partial
(x,H_1,\cdots,H_{n-1})}{\partial(x_1,\cdots,x_n)}
\end{eqnarray*}
where $\epsilon_{ijk\cdots l}$ denotes the Levi-Civita tensor. In
1994 Takthtajan \cite{takhtajan94ofo} made a geometric formulation
of Nambu-Poisson bracket, which satisfies the skew symmetry,
Leibniz rule and the fundamental identity, which is a
generalization of the Jacobi identity. According to
\cite{takhtajan94ofo}, the dynamics on the Nambu-Poisson manifold
is determined by $n-1$ Hamiltonians $H_1,\ldots,H_{n-1}$ and  is
described by the generalized Nambu-Poisson equation:
$$
\frac{dx}{dt} = \{H_1,\ldots,H_{n-1},x\} \label{NAM}
$$
The fundamental identity is given the $n=3$ as
\begin{eqnarray*}
\{\{G,H,F_1\},F_2,F_3\}  &+ & \{F_1,\{G,H,F_2\},F_3\} + \{F_1,F_2,\{G,H,F_1,F_3\}\} \\
& = & \{G,H,\{F_1,F_2,F_3\}.
\end{eqnarray*}
A function $F$ is called an integral of motion for the
Nambu-Hamilton system if its Nambu bracket with the Hamiltonians
$H_1,\ldots,H_{n-1}$ vanishes. A Nambu bracket of n integrals of
motion is again an integral of motion, this shows  that many
integrable systems possess Nambu structure.

\noindent{\bf Example 11: Lagrange system:\/}\\ The so called
Lagrange system \cite{takhtajan94ofo} which occurs in $SU(2)$
monopoles on ${\Bbb R}^3$
$$
\dot{x}_1 =  x_2x_3,\quad
\dot{x}_2  = x_3x_1,\quad
\dot{x}_3  =  x_1x_2
$$
can be written in the Nambu form
$$
\dot{x}_i = \{ x,H_1,H_2\}, \qquad H_1 = \frac{1}{2}(x_1^2 - x_2^2),\; H_2=\frac{1}{2}(x_1^2 -x_3^2).
$$
It turned out that many physical systems can be written in Nambu
form; like the Halpen system \cite{takhtajan94ofo}, some vortex
equations in fluid dynamics  \cite{baleanu98npf}, some two
dimensional incompressible flow equations \cite{pandit98ogn},
$SU(n)$ isotropic harmonic oscillators and  the $SO(4)$ Kepler
problem \cite{chatterjee96dsa}. The three dimensional
Lotka-Volterra equations (\ref{VOL}) also can be written in Nambu
form
$$
\dot{x} = M(x) \nabla H_1 \times \nabla H_2
$$
where $M(x) = x_1x_2x_3$ with the Hamiltonians $H_1$ and $H_2$ in (\ref{VOLH}).

Poisson systems can also be considered as a special case of a discrete gradient system \cite{mclachlan98uat}
\begin{equation}
\dot{x} = S(x)\nabla I(x) \label{DG}
\end{equation}
where $S(x)$ is skew-symmetric and $I(x)$ is a first integral. The
integral preserving methods based on the construction of discrete
gradients in \cite{mclachlan99giu, quispel96dgm} are similar to
those conserving schemes for Hamiltonian systems. The discrete
gradient method consists of the replacement of the  continuous
gradient $\nabla I(x)$ by  the discrete gradient
$\bar{\nabla}I(x)(x,\hat{x})$ which satisfies
\begin{eqnarray*}
\bar{\nabla}I(x)(x,\hat{x}) (\hat{x} - x)  & = & V(\hat{x}) - I(x), \\
\bar{\nabla}I(x)(x,\hat{x}) & = & \nabla I(x) + O(\hat{x} - x).
\end{eqnarray*}
The resulting discrete gradient system can then be constructed as
$$
\frac{\hat{x} - x}{\Delta t} = \tilde{S} \bar{\nabla}I(x)(x,\hat{x})
$$
where $\tilde{S}$ is any consistent skew-symmetric matrix such
that $ \tilde{S}(x,\hat{x}) = S\left (\frac{x +
\hat{x}}{2}\right)$.   It can be easily seen from the
skew-symmetry of $\tilde{S}$, that the first integrals are
preserved, i.e. $I(x) = I(\hat{x})$. Discrete gradients are not
unique; several examples of them are given in
\cite{mclachlan99giu, quispel96dgm}.

Systems with multiple integrals $I_1(x),\cdots,I_{n-1}(x)$  like
the Nambu systems (\ref{NAM}) can be written in the multi-gradient
form
$$
\dot{x} = S(x) \nabla I_1(x)\cdots\nabla I_{n-1}(x)
$$
where $S(x)$ is an $n$ skew-symmetric tensor. Discrete gradient systems which preserve the multiple gradients can be constructed
similarly \cite{mclachlan98uat}.

\section{Acknowledgements}
The author  acknowledges the support of the Swiss National Science
Foundation and is grateful to Ernst Hairer and Gerhard Wanner for
their hospitality during his stay at Universit\'{e} de Gen\`{e}ve,
in 2000. The author thanks also the referee for the helpful
comments.
\bibliographystyle{plain}

\begin{thebibliography}{abc}
\bibitem{ablowitz01dis} M.J. Ablowitz, B.M. Herbst and C.M. Schober,
Discretizations, integrable systems and computation, {\em J. Phys.
A: Math. Gen.\/} {\bf 34\/}, 10671-10693 (2001).
\bibitem{ablowitz96otn1} M.J. Ablowitz, B.M. Herbst and C.M. Schober,
On the numerical solution of the {S}ine-{G}ordon equation, {I}.
integrable discretizations and homoclinic manifolds, {\em Journal
of Computational Physics\/}, {\bf 126\/}, 299-314 (1996).
\bibitem{antonowicz90hso} M. Antonowicz and A.P. Fordy, Hamiltonian structure of nonlinear evolution equations,
In {\em Soliton Theory: A Survey of Results\/}, Manchester Univ.
Press, Manchester, (Edited by  A.P. Fordy), 273-312, (1990).
\bibitem{arnold98tmi} V.I. Arnold and B.A. Khesin, {\em Topological Methods in Hydrodynamics\/}, Springer, New York, (1998).
\bibitem{austin93api} M. Austin, P. S. Krishnaprasad and L.-S.
Wang, Almost {P}oisson integration of rigid rody systems, {\em
Journal of Computational Physics\/}, {\bf 107\/}, 105-117 (1993).
\bibitem{baleanu98npf} D. Baleanu,  and N. Makhaldiani, Nambu-{P}oisson reformulation of the finite-dimensional
dynamical systems, Ob ed. Inst. Yadernykh Issled., Dubna,(1998).
\bibitem{benzel93eco} S. Benzel, Z. Ge and C. Scovel, {E}lementary construction of higher order {L}ie-{P}oisson
integrators, {\em Physics Letters A\/}, {\bf 174\/}, 229-232
(1993).
\bibitem{black99aon} W. Black, J.A.C. Weideman, and B.M. Herbst, Comment: ``{A} note on an integrable
discretization of the nonlinear {S}chr\"odinger equation'', Y.
{B}. {S}uris,{\em Inverse Problems\/}, {\bf 15\/}, 807-810,
(1999).
\bibitem{blaszak98mto} M. B{\l}aszak, {\em Multi-{H}amiltonian Theory of Dynamical Systems\/},
Springer, Berlin, (1998)
\bibitem{budd01gia} C. Budd and M.D. Piggott, Geometric integration and its appliacations,
{\em Lecture notes for SCICADE01\/},  Dept. of Mathematical
Sciences, University of Bath, (2001).
\bibitem{buss00aae} R.S. Buss, Accurate and efficient simulation of rigid-body
rotations, {\em Journal of Computational Physics\/}, {\bf 164},
377-406 (2000).
\bibitem{channel91ifl} P. J. Channel and J. S. Scovel, {I}ntegrators for {L}ie-{P}oisson dynamical systems,
{\em Physica D\/}, {\bf 50\/}, 80-88 (1991).
\bibitem{channell90sio} P.J. Channell and C. S. Scovel, Symplectic integration of {H}amiltonian systems,
{\em Nonlinearity\/}, {\bf 3\/}, 231-259 (1990).
\bibitem{chatterjee96dsa} R. Chatterjee, Dynamical symmetries and {N}ambu mechanics,
{\em Letters in Mathematical Physics\/}, {\bf 36}, 117-126 (1996).
\bibitem{cherdantsev96lms} I. Cherdantsev and R. Yamilov,
Local master symmetries of differential-difference equations, In
{\em Symmetries and Integrability of Difference Equations\/},
American Mathematical Society, 51-61, (1996).
\bibitem{damianou00mhs} P.A. Damianou, Multiple Hamiltonian Structures for {T}oda systems of type
{A}-{B}-{C}, {\em Regular and Chaotic Dynamics\/}, {\bf 5\/},
17-32 (2000).
\bibitem{dubrovin99fma} B. Dubrovin, Frobenius manifolds and differential geometry of integrable hierarchies,
International School for Advanced Studies, ICTP, (1999).
\bibitem{ergenc96rkc} T. Ergen{\c{c}} and B. Karas{\"o}zen, Runge--{K}utta collocation methods for rigid
body {L}ie-{P}oisson equations, {\em International Journal
Computer Mathematics\/}, {\bf 62\/}, 63-71 (1996).
\bibitem{faddeev87hmi} L.D. Faddeev, and L.A. Takhtajan, {\em Hamiltonian Methods in the Theory of Solitons},
Springer, Berlin, (1987).
\bibitem{feng91sds} K. Feng and D.L. Wang, Symplectic difference schemes for {H}amiltonian systems in
general symplectic structure, {\em Journal of Computational
Mathematics\/}, {\bf 9\/}, 86-96 (1991).
\bibitem{frank97gif} J. Frank, W. Huang, B. Leimkuhler, Geometric integrators for classical spin systems,
{\em Journal of Computational Physics\/}, {\bf 133\/}, 160-172
(1997).
\bibitem{ge91esd} Z. Ge, {E}quivariant symplectic difference schemes and generating functions, {\em Physica
D\/}, {\bf 49\/}, 376-386 (1991).
\bibitem{ge90gfh} Z. Ge, Generating functions, {H}amilton-{J}acobi equations and symplectic groupoids on
{P}oisson manifolds, {\em Indiana University Mathematics Journal},
{\bf 39\/}, 859-876 (1990).
\bibitem{ge-zhong88lph}
Z. Ge and J. Marsden, {L}ie-{P}oisson {H}amiltonian-{J}acobi
theory and {L}ie-{P}oisson integrators, {\em Physics Letters A\/},
{\bf 133\/}, 134-139 (1988).
\bibitem{gumral93pso} H. G{\"u}mral and Y. Nutku, Poisson structure of dynamical systems with three
degrees of freedom, {\em Journal of Mathematical Physics\/}, {\bf
34\/},  5691-5723 (1993).
\bibitem{hairer00gi} E. Hairer,  C. Lubich and G. Wanner, {\em Geometric Numerical
Integration\/}, Springer, Berlin, (2002).
\bibitem{hairer00sif} E. Hairer and C.M. Schober,
Corrigendum to "Symplectic integrators for the {A}blowitz-{L}adik
discrete nonlinear {S}ch r\"odinger equation: implementation
issues[Phys. Lett. A {259} (1999),  140--151] , {\em Physics
Letters A\/}, {\bf 272\/}, 421-422 (2000).
\bibitem{islas01gif}
A.L. Islas, D.A. Karpeev and C.M. Schober, Geometric integrators
for the nonlinear {S}chr\"odingeringer equation, {\em Journal of
Computational Physics\/}, {\bf 173\/}, 116-148 (2001)
\bibitem{karasozen96cif} B. Karas{\"o}zen, Composite integrators for {B}i-{H}amiltonian systems,
{\em Computer \& Mathematics with Applications\/}, {\bf 32\/},
79-86 (1996).
\bibitem{karasozen00spf} B. Karas\"{o}zen, Structure preserving integrators for Volterra lattice equations,
Department of Mathematics, Middle East Technical University,
Ankara, (2000)
\bibitem{li95anf} S. T. Li and M. Qin, Lie-{P}oisson integration for rigid body dynamics,
{\em Computer \& Mathematics with Applications\/}, {\bf 30\/},
105-118 (1995). \bibitem{magri96eli} F. Magri,  Eight lectures on
integrable systems, In {\em Integrability of Nonlinear Systems\/},
Springer, 256-296, (1996).
\bibitem{mardsen99mgv} J.E. Marsden, G.P. Patrick, S. Shkoller,
Multi-symplectic geometry, variational integrators, and nonlinear
pde's, {\em Communications in Mathematical Physics\/}, {\bf
199\/}, 351-395 (1999).
\bibitem{mclachlan93elp} R. I. McLachlan, Explicit {L}ie-{P}oisson integration and the
{E}uler equations, {\em Physical Review E\/}, {\bf 71\/},
3043-3046 (1993).
\bibitem{mclachlan95cop} R. I. McLachlan, Comment on "{P}oisson schemes for {H}amiltonian
systems on {P}oisson manifolds", {\em Computer \& Mathematics with
Applications\/}, {\bf 29\/}, 1 (1995).
\bibitem{mclachlan98uat} R. I. McLachlan,  G. R. W. Quispel and N. Robidoux, A unified approach to {H}amiltonian systems,
{P}oisson systems, gradient systems, and systems with {L}yapunov
functions and/or first integrals, {\em Physical Review E\/}, {\bf
81\/}, 2399-2403 (1998).
\bibitem{mclachlan99giu} R. I. McLachlan,  G. R. W. Quispel and N. Robidoux, Geometric integration using discrete gradients,
{\em  Progress in Theoretical Physics\/}, {\bf 357\/}, 1021-1046
(1999).
\bibitem{nambu73ghd} Y. Nambu, Generalized {H}amiltonian dynamics, {\em Physical Review
D\/}, {\bf 7\/}, 2405-2412 (1973)
\bibitem{oevel89mam} W. Oevel, H. Zhang  and B.Fuchssteiner, Mastersymmetries and multi-{H}amiltonian formulations for some
integrable lattice systems, {\em Progress of Theoretical
Physics\/}, {\bf 81\/}, 294-308 (1989).
\bibitem{olver90cfa} P.J. Olver, Canonical forms and integrability of bi-{H}amiltonian systems,
{\em Physics Letters A\/}, {\bf 148\/}, 177-187 (1990).
\bibitem{olver93aol} P. J. Olver, {\em Applications of {L}ie Groups to Differential Equations\/},
Second Edition, Springer, New York, (1993).
\bibitem{olver93cff} P.J. Olver, Canonical forms for bi-{H}amiltonian systems,
In {\em Integrable systems (Luminy, 1991)\/}, Birkh\"auser,
Boston, 239-249, (1993).
\bibitem{pandit98ogn} S.A. Pandit and A.D. Gangal, On generalized {N}ambu mechanics,
{\em Journal of Physics A. Mathematical and General\/}, {\bf
31\/}, 2899-2912 (1998).
\bibitem{perelomov95sto} A. M. Perelomov, {\em Selected Topics on Classical Integrable Systems\/}, Troisieme Cycle de
la Physique, Universit\'{e} de Gen\`{e}ve, (1995).
\bibitem{perelomov90iso} A.M. Perelomov, {\em Integrable systems of classical mechanics and {L}ie
algebras\/}, Birkh\"auser, Basel, (1990).
\bibitem{puta99cps} M.Puta, I. Ca{\c{s}}u and A. Voitecovici, Constant
{P}oisson structures and the weighted {E}uler integrator, {\em
Journal of Computational and Applied Mathematics\/}, {\bf 111},
147-152 (1999)
\bibitem{quispel96dgm} G.R.W. Quispel and G.S. Turner, Discrete gradient methods for solving {O}{D}{E}s numerically
while preserving a first integral, {\em Journal of Physics A\/},
{\bf 29\/}, L341-L349 (1996).
\bibitem{rauch90hsa} S. Rauch-Wojciechowski, Hamiltonian structures and complete integrability in analytical
mechanics, In {\em Soliton Theory: A Survey of Results\/},
Manchester Univ. Press, Manchester, (Edited by  A.P. Fordy),
235-272, (1990).
\bibitem{reich94mcs} S. Reich, {M}omentum conserving symplectic
integrators, {\em Physica D\/}, {\bf 76\/}, 375-383, (1994).
\bibitem{reich00mrk} S. Reich, S, Multi-symplectic {R}unge-{K}utta collocation methods for
{H}amiltonian wave equations, {\em Journal of Computational
Physics\/}, {\bf 157\/}, 473-499, (2000)
\bibitem{sanz-serna94nhp} J.M. {S}anz{--}{S}erna and {M}.{P}. {C}alvo, {\em {N}umerical {H}amiltonian {P}roblems\/},
{C}hapman {\&} {H}all, (1994).
\bibitem{silva99gmf} A.C. da Silva and  A. Weinstein, {\em Geometric Models for Noncommutative
Algebras\/},American Mathematical Society, (1999).
\bibitem{schober99sif}
C.M. Schober, Symplectic integrators for the {A}blowitz-{L}adik
discrete nonlinear {S}chr\"odinger equation,  {\em Physics Letters
A\/}, {\bf 259\/}, 140-151 (1999).
\bibitem{suris97ano} Y.B. Suris, A note on an integrable discretization of the nonlinear
{S}chr\"odinger equation, {\em Inverse Problems\/},  {\bf 13\/},
1211-1136 (1997).
\bibitem{suris99idf}
Y.B. Suris, Integrable discretizations for lattice system: local
equations of motion and their {H}amiltonian properties, {\em
Reviews in Mathematical Physics\/}, {\bf 11\/}, 727-822 (1999).
\bibitem{takhtajan94ofo} L. Takhtajan, On foundation of the
generalized {N}ambu mechanics, {\em Communications in Mathematical
Physics\/}, {\bf 160}, 295-315 (1994).
\bibitem{tang97smf} Y. F. Tang,  L. V{\'a}zquez,  F. Zhang and
V. M. P{\'e}rez-Carc{\'i}a,{S}ymplectic methods for the nonlinear
{S}chr{\"o}dinger equation, {\em Computer \& Mathematics with
Applications\/}, {\bf 32\/}, 73-83 (1996).
\bibitem{yoshida90coh} H. {Y}oshida, {C}onstruction of higher order symplectic integrators, {\em Physics Letters A\/},
{\bf 150\/}, 262-268 (1990).
\bibitem{zeitlin91fao} V. Zeitlin, Finite-Mode analogs of 2{D} ideal hydrodynamics:
coadjoint orbits and local canonical structure, {\em Physica D\/},
{\bf 49\/}, 353-362 (1991).
\bibitem{zhu94psf} W. Zhu and M. Qin, Poisson schemes for {H}amiltonian systems on {P}oisson manifolds, {\em Computer \& Mathematics with
Applications\/}, {\bf 27\/}, 7-16 (1994).
\end{thebibliography}

\end{document}